\documentstyle[12pt]{article}

\newcommand{\mrm}[1]{\mathrm{#1}}

\newcommand{\alphaem}{\alpha_{\mrm{em}}}

\renewcommand{\d}{\mrm{d}}
\newcommand{\e}{\mrm{e}}

\newcommand{\g}{\mrm{g}}

\newcommand{\p}{\mrm{p}}
\newcommand{\q}{\mrm{q}}

\newcommand{\qbar}{\overline{\mrm{q}}}

\newcommand{\ee}{\e^+\e^-}

\newcommand{\qqbar}{\q\qbar}

{\end{list}}
\newcounter{enumct}

\newlength{\captivewidth}
\newlength{\abstwidth}
\def\thebibliographys#1{\subsection*{References}\list
  {[\arabic{enumi}]}{\settowidth\labelwidth{[#1]}\leftmargin\labelwidth
    \advance\leftmargin\labelsep
    \usecounter{enumi}}
    \def\newblock{\hskip .11em plus .33em minus -.07em}
    \sloppy
    \sfcode`\.=1000\relax}

\begin{document}

\sloppy

\pagestyle{empty}

\begin{flushright}
CERN--TH/95--153
\end{flushright}

\vspace{\fill}

\begin{center}
{\LARGE\bf On the non-perturbative part of}\\[4mm]
{\LARGE\bf the photon structure function$^a$}\\[10mm]
{\Large Gerhard A. Schuler} \\[3mm]
{\it Theory Division, CERN} \\[1mm]
{\it CH-1211 Geneva 23, Switzerland}\\[2mm]
{and}\\[2mm]
{\it Institut f\"ur Theoretische Physik, Universit\"at Regensburg}\\[1mm]
{\it D-93053 Regensburg, Germany}\\[1mm]
\end{center}

\vspace{\fill}

\begin{center}
{\bf Abstract}\\[2ex]
\begin{minipage}{\abstwidth}
We discuss a dispersion relation in the photon mass and show how
(in principle) model-independent constraints on the parton distribution
functions of the photon, notably a momentum sumrule, can be obtained.
We present two sets of parametrizations, SaS~1 and~2, corresponding
to two rather extreme realizations of the non-perturbative part.
Inclusive electron scattering off a real photon is found to be insufficient
to constrain the non-perturbative components. The additional
sensitivity provided by the photon virtuality is outlined.
Previous approaches to model the non-perturbative input distributions
are commented upon.
\end{minipage}
\end{center}

\vspace{\fill}
\noindent
\rule{6cm}{0.4mm}

\vspace{1mm} \noindent
{\Large {\bf $^a$}} Talk presented at the Xth Workshop on Photon--Photon
Collisions, Photon '95, Sheffield, England, 8--13 June 1995

\vspace{\fill}

\noindent
CERN--TH/95--153\\
June 1995

\clearpage
\pagestyle{plain}
\setcounter{page}{1}

\subsubsection*{1.~~Introduction}
Perturbative QCD predicts only the $Q^2$ evolution of the
parton distribution functions (PDFs)
of the photon $f_i^{\gamma}(x,Q^2)$ via a set of inhomogeneous
differential equations of the first kind. Hence the solutions
$f_i^{\gamma}(x,Q^2)$ require the specification of the PDFs
at some $Q^2 = Q_0^2$. Two ways exist to determine these non-perturbative
input distributions $f_i^{\gamma}(x,Q_0^2)$. The first one is
analogous to the determination of hadronic PDFs: At $Q_0$
large enough to be safely within the perturbative
regime ($Q_0 \sim 2\,$GeV),
the parameters of the input distributions
$f_a^{\gamma}(x,Q_0^2)$ (shapes and normalizations)
are fitted \cite{LAC} to the experimentally
measured distributions involving the
PDFs of the photon, which thus far means
to the available $F_2^{\gamma}(x,Q^2)$ data.
Since these data are currently restricted to large $x$,
only the $u$-valence distribution is known with some confidence.
In particular, basically no constraint on the gluon distribution
of the photon exists today in such an approach.

In the second approach one pretends to know the input distributions
at some very low scale $Q_0 \sim 0.5\,$GeV apart from a single, adjustable
parameter. The expectation is that, at such low scales,
the photon should essentially behave like a hadron and, correspondingly,
the PDF of the photon could be identified with appropriate hadronic ones.
The experimental evidence for this ansatz is, however, rather weak:
the only data for $Q^2$ below $2\,$GeV$^2$ come from the TPC/2$\gamma$
measurement \cite{TPC} of $F_2^{\gamma}(x,Q^2)$ (at an average $Q^2$ of about
$0.7\,$GeV$^2$) and consist of no more than a handful of points
in a limited $x$-range; the largest $x$-bins are moreover plagued by
resonance contributions. Nonetheless, basically all
recent experimental analyses
accept the hadron-like parametrization of $F_2^{\gamma}(x,Q_0^2)$
of TPC/2$\gamma$ as the non-perturbative input. The scale $Q_0$ is
considered as a free parameter and fitted to their data, hence
disregarding most of the potential of their own
$F_2^{\gamma}(x,Q^2)$ data to extract the non-perturbative part.
Rather, these analyses merely quantify
how compatible the more recent data are with the TPC/2$\gamma$ ansatz.
The actual, fitted value of $Q_0$ is, in fact, not a significant number
since it is strongly correlated with the size (and shape) of the
assumed non-perturbative input.

This correlation
is most easily seen by decomposing the PDFs of the
(real, i.e.\ $P^2=0$) photon as follows
\begin{equation}
f_a^{\gamma}(x,Q^2) -  f_a^{\gamma,\mrm{dir}}(x,Q^2)
  = f_a^{\gamma,\mrm{PT}}(x,Q^2,Q_0^2)
   + f_a^{\gamma,\mrm{NP}}(x,Q^2,Q_0^2)
\ ,
\label{generaldeco}
\end{equation}
where the second term on the LHS describes the
(properly normalized $Z_3 = 1 + O(\alphaem)$)
probability distribution of a photon to remain a photon
\begin{equation}
f_a^{\gamma,\mrm{dir}}(x,Q^2) = Z_3 \, \delta_{a\gamma} \,
\delta (1-x)
\ .
\label{dirgammaPDF}
\end{equation}
Being the solution of an inhomogeneous evolution equation,
the PDF of the photon can always be written as the sum of two terms,
as in (\ref{generaldeco}) where the first term on the RHS
is a particular solution of the {\em inhomogeneous}
equation with the boundary condition
\begin{equation}
f_a^{\gamma,\mrm{PT}}(x,Q_0^2,Q_0^2) = 0
\ .
\label{BCanom}
\end{equation}
The second term is a general solution of the corresponding
{\em homogeneous} evolution equation and
needs a (non-perturbative) input distribution at $Q^2=Q_0^2$:
\begin{equation}
f_a^{\gamma,\mrm{NP}}(x,Q_0^2,Q_0^2) =  \tilde{f}_a^{\mrm{NP}}(x)
\ .
\label{BCNP}
\end{equation}

At this point it should also be pointed out that the perturbative
$Q^2$ evolution is not treated correctly in
the experimental analyses of the photon structure function.
The scale $Q_0$ is fitted to the respective data on
\begin{eqnarray}
F_2^{\gamma}(x,Q^2) & = & 2 x \sum_{q} e_q^2 \left\{
f_q^{\gamma,\mrm{PT}}(x,Q^2,Q_0^2) + f_q^{\gamma,\mrm{NP}}(x,Q^2,Q_0^2)
\right\}
\nonumber\\
 & \equiv &  F_2^{\gamma,\mrm{PT}}(x,Q^2,Q_0^2) +
F_2^{\gamma,\mrm{NP}}(x,Q^2,Q_0^2)
\ ,
\label{F2decompo}
\end{eqnarray}
using the FKP parametrization \cite{FKP}
of $F_2^{\gamma,\mrm{PT}}(x,Q^2,Q_0^2)$. An error in the
$Q^2$ evolution of $F_2^\gamma(x,Q^2)$ arises because the hadronic part is
not evolved with $Q^2$ but kept fixed at the input scale,
$F_2^{\gamma,\mrm{NP}}(x,Q^2,Q_0^2) = F_2^{\gamma}(x)[\mrm{TPC}/2\gamma]$.
There is yet another error:
the FKP parametrization is based on a valence approximation,
and hence fails for $x < 0.3$ \cite{ftwoga}.

The second approach, namely approximating the photonic input distributions
at some low $Q_0\sim 0.5\,$GeV by hadronic ones, has also been
pursued in theoretical analyses \cite{GRV}.
Here the input distributions are identified with those
of the pion and $Q_0$ is fixed by theoretical prejudice. In order to
have an adjustable parameter, the overall normalization of the
input distributions is allowed to vary.
(This ``$K$-factor" is actually fitted to high-$Q^2$
($Q^2 \gg Q_0^2$) data only.
In this way one ``only" assumes the leading-twist formula to
evolve  perturbatively down to low scales but not to describe all
the physics at low scales.)

It should be stressed that the estimation of the
input distributions $f_a^{\gamma,\mrm{NP}}(x,Q_0^2,Q_0^2)$ by the ones
of the pion using vector-meson dominance (VMD) and the
additive quark model involves quite severe assumptions.
The PDFs of the $\qqbar$ ``bound states" of the photon
need not be the same as those of real vector mesons.
Moreover, the PDFs of the short-lived $\rho^0$-meson
may well differ in shape from those of the long-lived pion.
In addition, not only the shapes may differ. Also the relative normalizations
of the various PDFs can be different, e.g.\ down-quark to strange-quark
distributions, valence to sea to gluon distributions. Finally,
implicit (but never justified) in this approach is the
assumption of a momentum sumrule
and a constraint on the number of valence quarks.

\subsubsection*{2.~~The dispersion relation in the photon mass}
It will be shown here how a dispersion relation in the photon mass
can be used to obtain (in principle) model-independent constraints
on the PDFs of the photon, namely a momentum sumrule, as well as
constraints on the valence distribution and the overall normalization.

The moments of the photonic PDFs ($g(n) = \int_0^1 {\d} x x^{n-1} g(x)$)
can be represented as a dispersion integral in the photon mass $\sigma^2$
($P^2$ is the photon virtuality) \cite{Bj}
\begin{equation}
  f_a^\gamma(n,Q^2,P^2) = \int_0^{\infty} \frac{{\d} \sigma^2}
  {\sigma^2 + P^2} \ \rho_a(n,Q^2,\sigma^2)
\ .
\label{dispersion}
\end{equation}
Rather than describing the dispersion integral as the difference between
a ``point-like" part (contribution from the upper limit) and a
``hadronic" part (contribution from the lower limit),
it is more natural to separate short-distance
and long-distance parts by a scale $Q_0$, since the weight function $\rho_a$
possesses the scaling-violation pattern typical of ordinary hadronic
PDFs \cite{Bj}. At large values of $\sigma^2$, the
$\gamma\rightarrow\qqbar$ transition can be calculated perturbatively.
At lower values of $\sigma^2$ one enters the resonance region:
non-perturbative Regge poles will contribute signalling the appearance
of $\qqbar$ bound states. A general ansatz for the weight function
is therefore ($\rho' \equiv {\d}\rho/{\d}\sigma^2$)
\begin{eqnarray}
 \rho_a(n,Q^2,\sigma^2) & = &
\sum_{V=V_0}^{V_m(Q_0)}\ A_a^V(n,Q^2)\
   \delta\left(1 - \frac{\sigma^2}{m_V^2}\right)
+\sum_{V=V_0}^{V_m(Q_0)}\ B_a^V(n,Q^2)\
   \delta'\left(1 - \frac{\sigma^2}{m_V^2}\right)
\nonumber\\
 & & \; +~ \Theta\left(\sigma^2 - Q_0^2\right)\, \left\{
  \alpha_a(n,Q^2,\sigma^2) +
      \beta'_a(n,Q^2,\sigma^2) \right\}
\ .
\label{generalansatz}
\end{eqnarray}

The coefficients $A_a^V$, $B_a^V$, $\alpha_a$ and $\beta_a$
can be determined as follows. VMD is known to well
describe photon-hadron interactions over a wide range of energies,
from $\sqrt{s}$ of a few GeV up to the HERA energy ($200\,$GeV). Hence,
to very good approximation, one may neglect $A_a^V$ and take
\begin{equation}
  B_a^V(n,Q^2,Q_0^2) = \left(\frac{e}{f_V}\right)^2\
  f_a^{\gamma,V}(n,Q^2,Q_0^2)
\ .
\end{equation}

In order to obtain $\alpha_a$ and $\beta_a$ one first notices
that for $Q_0^2 \ll P^2 \ll Q^2$ the resonance contributions to
(\ref{generalansatz}) are suppressed. Moreover, in this limit
the PDF of a virtual photon (i.e.\ the LHS of (\ref{dispersion}))
can be calculated within perturbative QCD \cite{Uematsu}.
Then one expresses these distributions as an integral of
``state" distribution functions $f_a^{\gamma,\q\qbar}(x,Q^2,\sigma^2)$:
\begin{equation}
 f_a^{\gamma}(x,Q^2,P^2) =
  \int_{P^2}^{Q^2} \frac{{\d} \sigma^2}{\sigma^2}\;
    \frac{\alphaem}{2\pi} \, \sum_{\q} 2 e_{\q}^2 \,
        f_a^{\gamma,\q\qbar}(x,Q^2,\sigma^2)
\ ,
\label{virtualPDF}
\end{equation}
which obey the standard, homogeneous evolution equations
with the boundary condition \cite{ftwoga}
\begin{equation}
f_a^{\gamma,\q\qbar}(x,\sigma^2,\sigma^2) = f_a^{\gamma,\q\qbar}(x)
 \equiv \frac{3}{2} \, \left( x^2 + (1-x)^2 \right) \,
( \delta_{a\q} + \delta_{a\qbar} )
\ .
\label{fxinit}
\end{equation}
Equation (\ref{virtualPDF}) yields an expression for the sum
$\alpha_a$ plus $\beta'_a$. The decomposition into $\alpha_a$ and
$\beta_a$ is more difficult. Generalized VMD arguments suggest
$\alpha_a \ll \beta_a$ and hence we
arrive at
the final expression for the PDFs of the virtual photon
\begin{eqnarray}
 f_a^{\gamma}(n,Q^2,P^2) & = &  f_a^{\gamma,\mrm{NP}}(n,Q^2,P^2)
       + f_a^{\gamma,\mrm{PT}}(n,Q^2,P^2)
\nonumber\\
 & \equiv &
  \sum_{V=V_0}^{V_m(Q_0)}\
   \left( \frac{m_V^2}{m_V^2 + P^2} \right)^2\
  \frac{4\pi\alphaem}{f_V^2}\
    f_a^{\gamma,V}(n,Q^2,Q_0^2)
\nonumber \\  & & \qquad
 + \int_{Q_0^2}^{Q^2}
  \frac{\sigma^2 \d \sigma^2}
   {\left(\sigma^2 + P^2 \right)^2}\
    \frac{\alphaem}{\pi}\
      \left( \sum_q e_q^2 \right)
        f_a^{\gamma,\q\qbar}(n,Q^2,\sigma^2)
\ .
\label{finalexp}
\end{eqnarray}
Equation (\ref{finalexp}) contains three constraints on the
non-perturbative distributions. The first two follow from the
fact that $f_a^{\gamma,V}$ are ordinary mesonic PDFs. Hence they
should respect the number of valence quarks
\begin{equation}
  1 = f_{\q,\mrm{val}}^{\gamma,V}(n=1,Q^2,Q_0^2) \equiv
     \int_0^1 {\d} x\ f_{\q,\mrm{val}}^{\gamma,V}(x,Q^2,Q_0^2)
\label{numbval}
\end{equation}
and obey the momentum sumrule
\begin{equation}
  1 = \sum_{a=\q,\qbar,\g} f_a^{\gamma,V}(n=2,Q^2,Q_0^2) \equiv
     \int_0^1 {\d} x\ x\ f_a^{\gamma,V}(x,Q^2,Q_0^2)
\ .
\label{momsumrule}
\end{equation}

The third constraint on the photonic PDFs follows
from the observation that the RHS of (\ref{finalexp})
has to be $Q_0$-independent. This fixes the overall normalization
and, in turn, gives a momentum sumrule for the photonic PDFs:
\begin{eqnarray}
  1 - Z_3 & \equiv & \sum_{a=\q,\qbar,\g}
   \int_0^1 {\d} x\ x\ f_a^{\gamma}(x,Q^2,P^2)
\nonumber\\
 & = &
  \sum_{V=V_0}^{V_m(Q_0)}\
   \left( \frac{m_V^2}{m_V^2 + P^2} \right)^2\
  \frac{4\pi\alphaem}{f_V^2}\
 + \int_{Q_0^2}^{Q^2}
  \frac{\sigma^2 \d \sigma^2}
   {\left(\sigma^2 + P^2 \right)^2}\
    \frac{\alphaem}{\pi}\
      \left( \sum_q e_q^2 \right)
\label{photonsumrule}
\end{eqnarray}
since the perturbative state distributions
$f_a^{\gamma,\q\qbar}(n,Q^2,\sigma^2)$ obey
a relation analogous to (\ref{momsumrule}).

This last equation is, in fact, nothing but the observation that
the vacuum fluctuations of the photons probed in $e\gamma$ interactions
are precisely the ones seen in $\ee \rightarrow$ hadrons. The
probability per unit $\sigma^2$ of their occurrence is given by the
cross section of the latter reaction
\begin{equation}
  1 - Z_3 = \int_0^{Q^2} \d \sigma^2 \left( \frac{\sigma^2}
    {\sigma^2 + P^2} \right)^2\ \frac{\sigma_{tot}(\ee \rightarrow
   X(\sigma) )}{4\pi^2\alphaem}
\ .
\label{eerelation}
\end{equation}
Using the narrow-width approximation for the resonance (low-mass)
contribution and the parton-model result at high masses
\begin{eqnarray}
\lefteqn{
  \sigma_{tot}(\ee \rightarrow X(\sigma)) = }
\nonumber \\ & &
\sum_{V} 4\pi^2 \alphaem
  \left( \frac{e}{f_V} \right)^2\ \delta\left( \sigma^2 - m_V^2 \right)
  + \Theta\left(\sigma^2 - Q_0^2\right) N_C \left( \sum_q e_q^2 \right)\
  \frac{4\pi\alphaem}{3 \sigma^2}
\label{sigtotee}
\end{eqnarray}
the result (\ref{photonsumrule}) is recovered.

Once $Q_0$ has been determined for a given number of vector mesons,
(\ref{photonsumrule}) tells us how $Q_0$ has to be changed
in order to compensate
the inclusion (or omission) of a vector meson. For a given number
of included vector mesons, the value of $Q_0$ can be determined from
the continuity requirement of the $\ee$ annihilation cross section or, say,
the total $\gamma\p$ cross section \cite{gammap}.

What remains undetermined in the approach sketched above is the
{\em shape} of the non-perturbative input distributions.
In line with the argument that hard processes
probe short time scales, the contributions from the various
vector mesons should be added coherently.
Also, to rather good approximation,
an $SU_3$-symmetric sea distribution $s(x)$
can certainly be assumed.
Then the non-perturbative input distributions $f_a^{\gamma,\mrm{NP}}
(x,Q_0^2,Q_0^2)$ are given in terms of three distributions,
a valence distribution $v(x)$, a gluon distribution $g(x)$, and
the sea distribution $s(x)$. These distributions can be determined
through a fit to the available (real photon) $F_2^\gamma(x,Q^2)$ data,
subjected to the constraints (\ref{finalexp}),
(\ref{numbval}) and (\ref{momsumrule}).

Two extreme scenarios can be considered.
In the first, VMD is restricted to the well-established
$\rho^0$, $\omega$, $\phi$ states. The scale $Q_0$ is then known to
be $Q_0 \approx 0.6\,$GeV from an analysis of the $\gamma\p$ total
cross section \cite{gammap}.
This ``low-$Q_0$" fit (SaS set~1 distribution functions;
for details of the fit see \cite{ftwoga})
is essentially a three-parameter fit, two for the
shape of the valence distribution and one for the normalization of the
sea. The shapes of both the gluon and sea distributions are hardly
constrained by current $F_2^\gamma$ data and take on values of an
educated guess. The main theoretical error of this scenario
arises from the use of perturbation theory down to
rather low values of $Q^2$.

The spirit of the second analysis is
opposite: take $Q_0$ well within the perturbative domain ($Q_0=2\,$GeV)
at the expense of parametrizing the effects of additional vector mesons
(besides $\rho^0$, $\omega$, and $\phi$)
by a simple factor $K$ to be fitted to the data
\begin{equation}
 \sum_{V=V_0}^{V_m(Q_0)}
    \frac{4\pi\alphaem}{f_{V}^2}
   f_a^{\gamma,V}(x) \approx  K(Q_0)\,
     \sum_{V=\rho^0,\omega,\phi}
    \frac{4\pi\alphaem}{f_{V}^2}
       {f}_a^{\gamma,V}(x)
\ .
\label{Kfactor}
\end{equation}
This ``high-$Q_0$" fit (SaS set~2 distribution functions
\cite{ftwoga}) contains two additional parameters compared to the
low-$Q_0$ fit, one parameter characterizing the necessary hard component
of the valence distributions at larger values of $Q_0$, and the value
of $K$.

Note that, if higher-twist effects and other uncertainties
were negligible, the following dependence of $K$ on $Q_0$ should hold
\begin{equation}
  K(Q'_0)
  =
  K(Q_0) + \frac{\sum_{\q} e_{\q}^2}
                   {\pi \sum_{V=\rho,\omega,\phi} 4\pi/f_V^2}
  \ln \frac{Q'^2_0}{Q_0^2}
  \approx
K(Q_0) + 0.770  \ln \frac{Q'_0}{Q_0}
\ .
\label{normrelation}
\end{equation}
With $K(0.6\,\mrm{GeV})=1$, (\ref{normrelation}) predicts
$K(2\,\mrm{GeV}) = 1.93$, to be compared with $2.42$, the
outcome of the high-$Q_0$ fit.
Two facts account for almost all the discrepancy. First,
the $\chi^2$ analysis suggests that $K(0.6\,\mrm{GeV})$ should
slightly exceed unity ($K=1.17$) increasing the prediction
(\ref{normrelation}) to $K(2\,\mrm{GeV}) = 2.10$. Second, the
high-$Q_0$ fit includes only data above $Q^2 = 4\,$GeV$^2$, while
the low-$Q_0$ fit includes data down to $0.71\,$GeV$^2$.
Indeed, good agreement is found if the latter fit is also restricted
to the high-$Q^2$ data. This indicates that
higher-twist contributions significantly affect $F_2^\gamma$
at low $Q^2$, but also
that the leading-twist evolution of PDFs is still valid
down to $Q_0 = 0.6\,$GeV.

Distributions involving the inclusive PDFs of the real photon are not
sufficient to disentangle the non-perturbative part.
The high-$Q_0$ and low-$Q_0$ SaS sets of PDFs describe the $F_2^\gamma$
data equally well \cite{ftwoga},
although the non-perturbative parts have very different
shapes and normalizations.
For example, the momentum fractions carried by the
perturbative and non-perturbative parts are
quite different for the two cases but not visible with a real photon
($P^2=0$) target:
\begin{eqnarray}
Q_0 =  2.0\,\mrm{GeV}:\quad
   1-Z_3 & = & \alphaem \left\{ 1.33 +
     \frac{\alphaem}{\pi} \left( \sum_q e_q^2 \right)\
    \ln \frac{Q^2}{4\,\mrm{GeV}^2} \right\}
\nonumber\\
Q_0 =  0.6\,\mrm{GeV}:\quad
    1-Z_3 & = & \alphaem \left\{ 0.55 +
     \frac{\alphaem}{\pi} \left( \sum_q e_q^2 \right)\
    \ln \frac{Q^2}{0.36\,\mrm{GeV}^2} \right\}
\nonumber\\
  & \approx &  \alphaem \left\{ 1.06 +
     \frac{\alphaem}{\pi} \left( \sum_q e_q^2 \right)\
    \ln \frac{Q^2}{4\,\mrm{GeV}^2} \right\}
\ .
\label{pzeromom}
\end{eqnarray}
Additional information can, and has to, be obtained from two sides:
perturbative and non-perturbative parts lead to differences in
the hadronic final states of photon-induced reactions \cite{gammap},
and also show a different dependence on the photon virtuality
as is evident, e.g.\ from (\ref{photonsumrule}).
Further studies, both theoretically and experimentally,
to exploit the sensitivity to the non-perturbative part are
highly desirable.

\vspace{\baselineskip}\noindent
\textbf{Acknowledgements:} I congratulate the organizers
for a stimulating and lively meeting, and warmly thank Torbj\"orn
Sj\"ostrand for an enjoyable collaboration.

\begin{thebibliographys}{99}

\bibitem{LAC}
M.\ Drees and K.\ Grassie,  Z.\ Phys.\ {\bf C28} (1985) 451;
\hfill\\
H.\ Abramowicz, K.\ Charchula and A.\ Levy,
Phys.\ Lett.\ {\bf B269} (1991) 458;
\hfill\\
K.\ Hagiwara, M.\ Tanaka, I.\ Watanabe and T.\ Izubuchi,
Phys.\ Rev.\ {\bf D51} (1995) 3197;
\hfill\\
L.E.\ Gordon and J.K.\ Storrow, Z.\ Phys.\ {\bf C56}
(1992) 307

\bibitem{TPC}
TPC/$2\gamma$ collab., H.\ Aihara et al., Z.\ Phys.\ {\bf C34} (1987) 1
 and Phys.\ Rev.\ Lett.\ {\bf 58} (1997) 97

\bibitem{FKP}
J.H.\ Field, F.\ Kapusta and L.\ Poggioli, Phys.\ Lett.\
{\bf B181} (1986) 362 and Z.\ Phys.\ {\bf C36} (1987) 121;
\hfill\\
F.\ Kapusta, Z.\ Phys.\ {\bf C42} (1989) 225

\bibitem{ftwoga}
G.A. Schuler and T. Sj\"ostrand, CERN--TH/95--62 and LU TP 95--6,
to appear in Z.\ Phys.\ {\bf C}

\bibitem{GRV}
M.\ Gl\"{u}ck, E.\ Reya and A.\ Vogt, Phys.\ Rev.\
{\bf D46} (1992) 1973;
\hfill\\
P.\ Aurenche, M.\ Fontannaz and J.-Ph.\ Guillet,
Z.\ Phys.\ {\bf C64} (1994) 621

\bibitem{Bj}
 J.D.\ Bjorken, ``Two topics in quantum chromodynamics",
 SLAC-PUB-5103, December 1989 (Summer Inst., Carg\`{e}se, 1989, p.\ 217)

\bibitem{Uematsu}
T.\ Uematsu and T.F.\ Walsh, Phys.\ Lett.\ {\bf B101} (1981) 263
and Nucl.\ Phys.\ {\bf B199} (1982) 93;
\hfill\\
G.\ Rossi, Univ.\ California at San Diego preprint UCSD-10P10-227
(1983) and Phys.\ Rev.\ {\bf D29} (1984) 852

\bibitem{gammap}
G.A. Schuler and T. Sj\"ostrand, Phys. Lett. {\bf B300} (1993) 169,
Nucl. Phys. {\bf B407} (1993) 539;
\hfill\\
T.\ Sj\"ostrand, these proceedings

\end{thebibliographys}

\end{document}